\documentclass{aa}  
\newcommand{\kms}{km\,s$^{-1}$\,}

\usepackage{graphicx}
\usepackage{epsfig,natbib,txfonts}

\usepackage{txfonts}
\begin{document}
   \title{Comparison between 2-D and 3-D codes in dynamical simulations of gas flow in  barred galaxies}

   \subtitle{}

   \author{I. P\'erez \inst{1}\thanks{Veni Fellow}\fnmsep\inst{2}\thanks{Associate Researcher}}

   \offprints{I. P\'erez}

\institute{Kapteyn Astronomical Institute, University of Groningen, the Netherlands
\\ email:isa@astro.rug.nl \and Departamento de F\'isica te\'orica  y del Cosmos,
 Universidad de Granada, Spain\
             }


 
  \abstract
{One of the ways to determine the contribution of the dark halo to the gravitational potential of a galaxy is the study of non-circular (streaming) motions and the associated gas shocks in the bar region.  These motions, determined by the potential in the inner parts, can break the disk-halo degeneracy. Here, two main
fluid dynamical approaches have been chosen to model the non-circular motions in the bar region; a 2-D Eulerian grid code
for an isothermal gas (FS2) and a 3-D smoothed particle hydrodynamic
code (N-body/SPH)} 
   {The aim of this paper is to compare and quantify the differences of the gas flows in rotating barred potential obtained using two different fluid dynamical approaches. We analyse the effect of using  2-D and a 3-D codes in the calculation of gas flow in barred galaxies and to which extend the results are affected by the code. To do this, we derive the velocity field and density maps for the mass model of NGC~4123 using a 3-D N-body/SPH code and compare the results to the previous 2-D Eulerian grid code results.}
   {Numerical modelling, 3-D N-body/SPH simulations}
   {The global velocity field and the gas distribution is very similar in both models. The study shows that the position and strength of the shocks
developed in the SPH simulations do not vary significantly compared to
the results derived from the 2-D FS2 code. The largest velocity difference across the shock is 20\kms between the 2-D and 3-D fluid dynamical models.}
   {The results obtained in the studies deriving the dark matter content of barred galaxies using the bar streaming motions and strength and position of shocks are robust to the fluid dynamical model used. The effect of 2-D and 3-D modelling can be neglected in this type of studies.}

               
   \maketitle
%

\section{Introduction}

There have been several studies addressing the distribution of dark
matter in galaxies using the non-axisymmetry of the potential of
barred galaxies: a) the fluid dynamics approach with the work
by~\citet[][hereafter WSW01]{weiner01} for the modelling of NGC~4123;
~\cite{perez3} and \cite[][hereafter PFF04]{perez4} for the study of
the dark matter content of 5 barred galaxies (NGC~5505, NGC~7483,
NGC~5728, and NGC~7267); and b) the sticky-particle approach
by~\cite{rautiainen} modelling the dynamics of  ESO~566-24,
and~\cite{salo} who modelled the H$\alpha$ velocity field of IC~4214
to derive the halo contribution.

The streaming motions and the associated gas shocks in the bar region
are determined by the potential in the inner parts of galaxies. The
velocity field of an axisymmetric galaxy does not allow to uniquely
disentangle the contribution of the halo from that of the disk.The
signatures of non-circular motions, however, can break the disk-halo
degeneracy and be used to obtain the contribution of the dark halo to
the potential. Common to all the modelling, the stellar potential is generated directly from the
broad band galaxy images with some recipe to treat the vertical distribution and some
assumptions about the mass-to-light (M/L) of the stellar
populations. The velocity fields obtained in this way are then
compared to the observed kinematic information to determine if they can
be reproduced by the models.

The main result, common to all the modelling carried out up to now, is
the fact that the gravitational field in the inner region is mostly
provided by the stellar luminous component. The bar pattern speeds
found by the different groups are consistent with fast rotators, and
the best fitting M/L ratios obtained are compatible with M/L ratios
derived from current population synthesis models. Although it is a very
powerful method, only a handful of galaxies have been modelled in this
way. 

Two main fluid dynamical approaches have
been chosen for this type of modelling: an Eulerian grid code for an
isothermal gas (FS2) and a smoothed particle hydrodynamic code
(N-body/SPH). SPH is a Lagrangian method for solving Euler's equation of motion. In this technique, the system is represented by a set of particles and the gas properties are calculated by a weighted average over the neighbouring particles.  On the other hand, the Eulerian grid is fixed in space and time,  the grid nodes and cells remain spatially fixed while the materials flow through the mesh. In the modelling carried out by WSW01, the scale height is introduced
as a smoothing length for the potential while the approach chosen by PFF04 uses
a full 3-D calculation of the forces. For future work and as a
consistency check, it is important to understand whether the results
obtained are dependent a) on the dynamical code used and b) 2D versus
3-D. No significant differences are expected between the SPH and the
Eulerian grid approach. Previous simulations~\cite[][hereafter PA00]{englmaier,patsis00}, running
2-D SPH under the same conditions as used by~\cite{athanassoula92}
with the FS2 Eulerian grid code indicated that the main features, such
as the strength of the shocks in the bar are quite similar and small
differences arise due to the statistical nature of SPH or to numerical
and artificial viscosities used in the different codes. Both studies pointed out the importance of certain parameters used in the simulations for the outcome of the modelling, such as the sound speed and the way the non-axisymmetric part of the potential is introduced.  The
use of a 2-D versus a 3-D code might introduce large variations in the
radial forces due to the different treatment of the vertical forces,
which in turn might vary the outcome of the analysis and the
conclusions derived from this methodology. In this paper, we
investigate the effect of 2-D vs 3-D numerical modelling in the gas
flows of rigidly rotating bar potentials. In order to test this effect, we have
modelled the gas flow in NGC~4123 using the mass
model and images provided by B.J. Weiner but using the 3-D N-body/SPH
code used in PFF04. The parameters chosen for the SPH simulations have
been taken from WSW01.  In future work we will address the comparison
between the sticky particle codes and the fluid dynamical methods. The
modelling is presented in Section~\ref{modelling}, the results and
comparison are presented in Section~\ref{comparison}. Finally,
discussion and conclusions are presented in Section~\ref{conclusions}

\section{Modelling}
\label{modelling}

The methodology for this study is simple, we derive the mass model for
NGC~4123 in a way similar to that of WSW01. We then run the SPH
models to obtain the gas distribution and velocity field, using the
best fit parameters (M/L and bar pattern speed) in WSW01. And finally, the gas
distribution and the velocity field are compared to the results
presented in WSW01. In this section we will present a brief
summary of both codes and initial conditions, emphasising the SPH
carried out in this work and explaining the way in which we derive the mass distribution.

The FS2 code is a second-order, flux splitting, Eulerian grid code for
an isothermal gas in an imposed gravitational potential, originally
written by G.D.~\cite{vanalbada}. The sound speed of the gas is 8\kms.The code does not include
self-gravity of the gas. The barred potential rotates at a fixed
pattern speed $\Omega_{B}$, and the grid rotates with the bar.  In WSW01, 0.1 Gyr are allowed to fully grow the bar in order to steadily adjust the flow. The
initial gas surface density is set to be 10M$_{\odot}$~pc$^{-2}$ inside a radius of 8 kpc; outside that radius
it falls off exponentially. A $sech^{2}(z/z_{0})$, where $z_{0}$
is the scale height, is assumed for the vertical distribution. With this
distribution, the accelerations are calculated in the midplane using a
Fourier transform method to convolve a Green's function with the
surface brightness distribution~\cite{hockney}. The scale height is
effectively a smoothing length for the potential. For a complete
explanation of the code and the initial conditions refer to WSW01.

The 3-D N-body and hydro code was initially developed by the Geneva
Observatory galactic dynamics group for spiral galaxy
studies~\citep{pfenniger93,fux97,fux99}.  The, initially
self-consistent, code was modified to use a fixed rotating
potential. The stellar potential is fixed using the observed light
distribution. The potential is calculated using a particle-mesh technique with cylindrical-polar grid and the short range forces are softened using a variable homogeneous ellipsoid kernel with principal axes matched to the local grid resolution. The pressure forces and viscous forces are derived by 3-D Smooth Particle Hydrodynamics, SPH~\citep{benz90,fux99}. The gas is taken to be isothermal with a sound speed of 10\kms and an adiabatic index of 5/3, corresponding to the atomic hydrogen. For more details about the code, refer to \cite{fux99}
or his Ph.D. Thesis, \cite{fux97}. For details about the code in
the form used for this work refer to PFF04.

The grid size used is 95 cells in the radial, 96 in the azimuthal and
1214 (including doubling up) in the vertical directions. The vertical
resolution is set to 0.05 times the scale-height adopted for the
luminous mass distribution. The number of gas particles used is 300
000. The barred potential rotates at fixed pattern speed $\Omega_{B}$
and there is no self-gravity.  These simulations were run; first, using
the Alpha Server Sc system at the Australian National University super
computer facility; and later, the Cray SV1e super computer at
Groningen University.

For the initial conditions see PFF04. The initial gas density
in the simulations consists of one component. The radial distribution
for the gas is a Beta function with a standard deviation set to the
scale-length of the visible disk and radially vanishing at a distance
4 times this scale-length. The scale-length assumed for NGC~4123 is 3.2~kpc, as derived from the $I$-band luminosity distribution in~\cite{weiner01B}. The vertical distribution of the gas is
generated directly by solving the hydrostatic equilibrium equation for
an isothermal gas. The gas particles have pure circular motion with
cylindrical rotation and zero velocity dispersion, since the effective
dispersion is taken into account in the pressure component of the
SPH. The bar is chosen to grow linearly during three bar rotations, so
that the gas flow can steadily adjust to the forcing bar without
requiring much CPU time. In previous simulations (PFF04)
different onset times were checked to ensure that at the chosen
onset time the particle flow adjusted steadily, finally setting the onset time to 3 bar rotations.

We have used the $I$-band light distribution, and exponential vertical
profile with a scale-height of 200~pc to derive the potential in a
similar fashion as in WSW01. They constructed a bi-symmetric image by
rotating the image 180$^{\circ}$ and averaging the rotated and
original images. The central point source was removed and modelled as
a spherical Gaussian distribution $\rho(r) \propto
exp(-r^{2}/2r^{2}_{\rm c})$ with a scale radius (r$_{\rm c}$) of 200~pc, refer to WSW01
for details about the mass distribution. Since the model used for
comparison is the maximum disk model, no dark matter halo profile was
included in the derivation of the mass. Because the potential in the
bar region is dominated by the stellar component we do not expect a
significant difference in the results when excluding this
component. The model parameters used were  M/L of 2.25 and pattern speed of 20
km~s$^{-1}$kpc$^{-1}$, corresponding to a $R_{\rm cor}$/R$_{\rm bar}$
= 1.35, where R$_{\rm cor}$ is the corotation radius and R$_{\rm bar}$
is the bar semi-major axis. These parameters correspond to the
best-fit model in WSW01. To derive the velocity field,
we used the parameters derived by \cite{weiner01B} from their
kinematic data; i.e. i = 45~$^{\circ}$ and major axis of the galaxy at
57~$^{\circ}$ North through West.

Both codes were run without gas self-gravity. Low resolution
simulations have shown that gas self-gravity is irrelevant for the results in
this work~\citep{perez3,perez4}.

The gas density distribution, for the SPH modelling, clearly settles into a stationary
configuration after 3 bar rotations, a nuclear ring develops
perpendicular to the bar, associated to the x$_{2}$ family of orbits .  From the frequency plot of the axisymmetric
case, we see that NGC4123 possesses two ILR for the derived pattern speed. The
presence of the ILR was already suggested  in~\cite{weiner01B} from
the offset of the dust lanes in the colour images of NGC~4123.

\section {Comparison and results}
\label{comparison}

\begin{figure*}
\begin{center}
\vspace{0cm}
\hspace{-1.5cm}\hbox{\epsfxsize=9.5cm \epsfbox{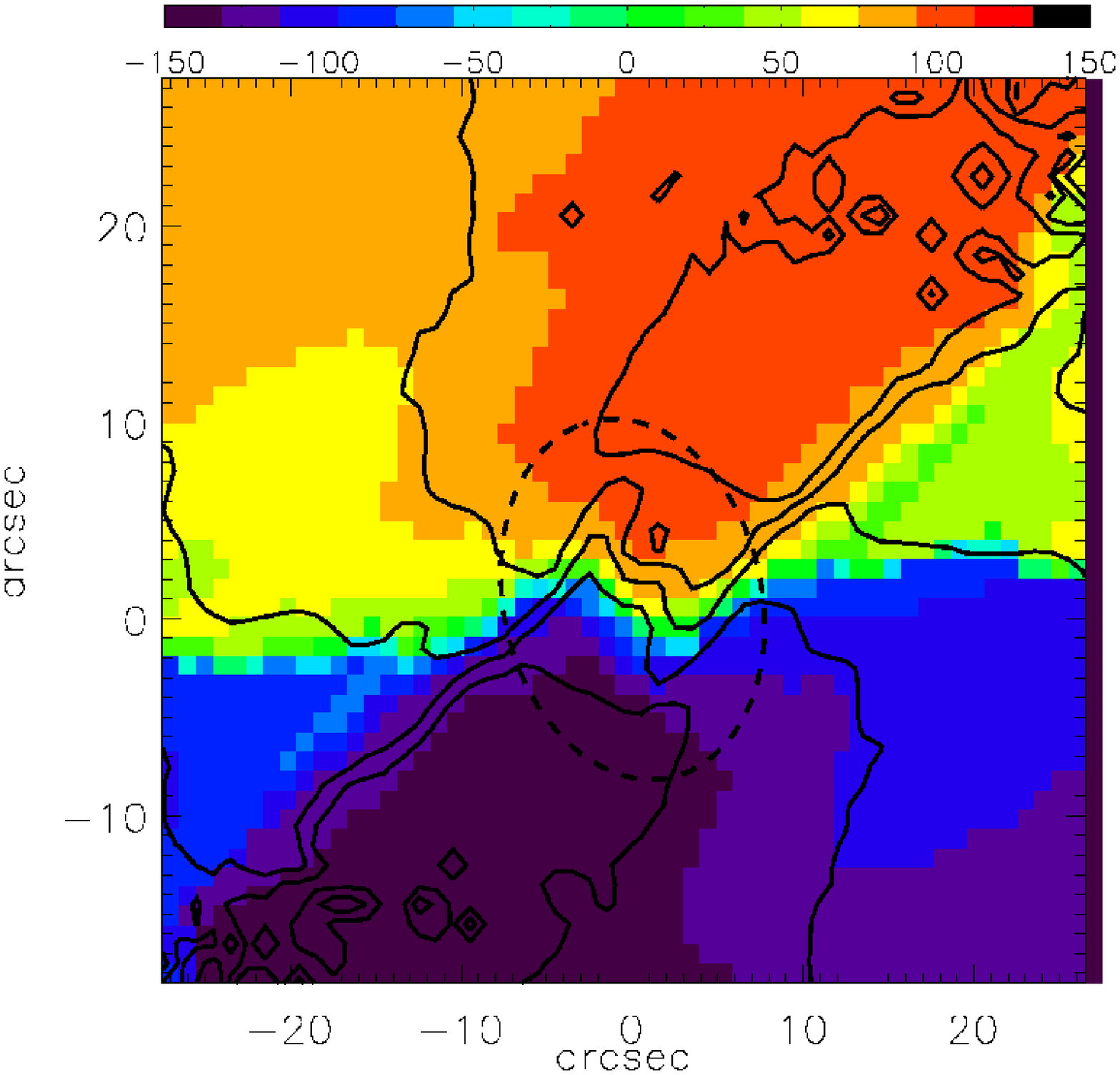}}
\hbox{\epsfxsize=9.5cm \epsfbox{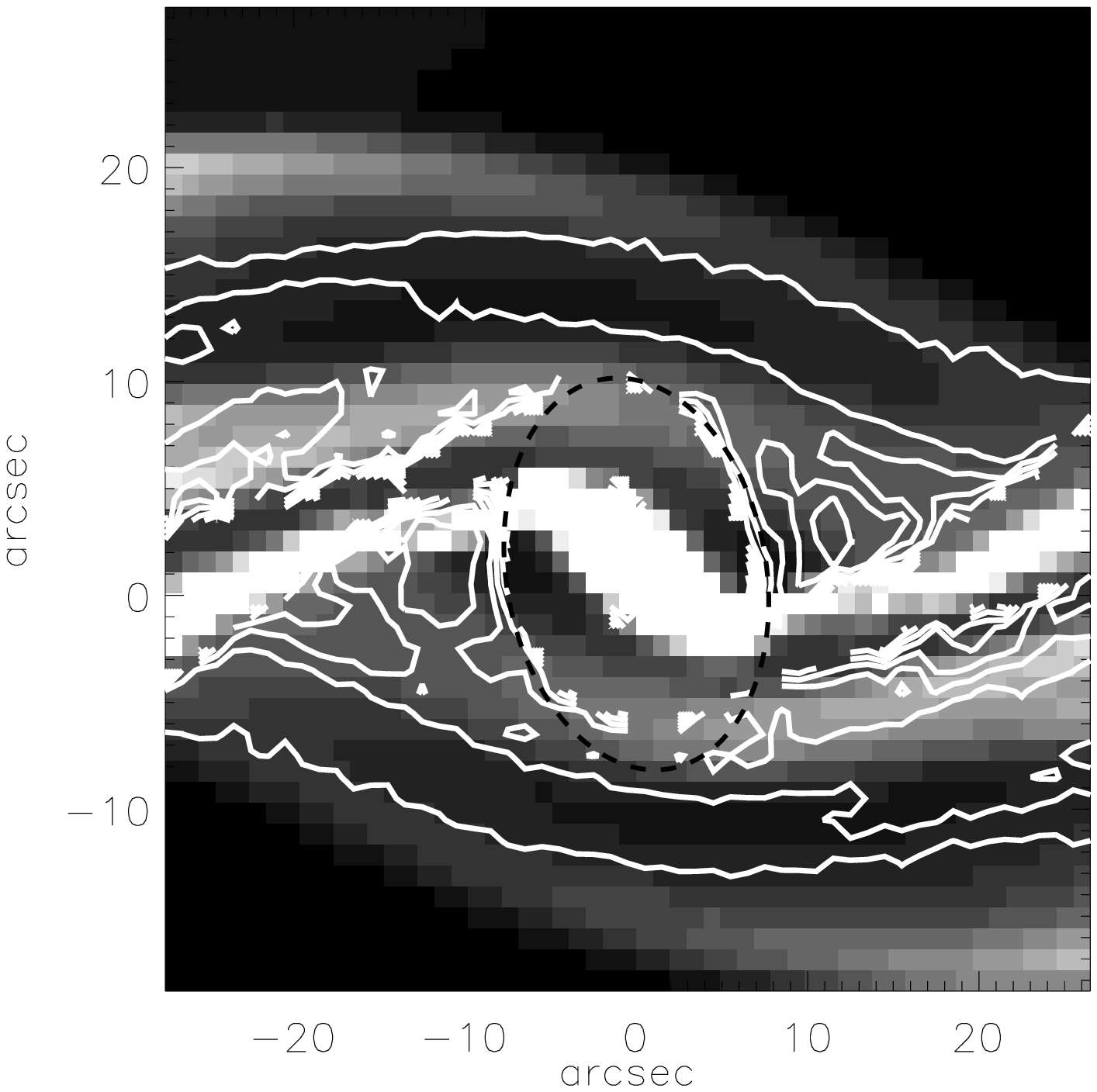}}
\caption{On the left panel; a) the velocity field contours in the bar region,
projected into the plane of the sky, using  i = 45~$^{\circ}$ and
major axis of the galaxy at 57~$^{\circ}$ northwest from WSW01. Superposed are
the isovelocity contours of the projected velocity field as derived
in this work, with levels corresponding to intervals of 30~km~s$^{-1}$. On the right panel; b) density distribution from WSW01, superposed are the isodensity contours from the particle distribution of the SPH for the mass distribution of NGC~4123
derived from the $I$-band after 4 bar rotations. The non-axisymmteric
component is fully grown after three bar rotations and the bar patter
speed is rotating clockwise in the inertial frame. Notice the agreement in the regions of high
density within the bar region.} 
\label{fig:velocity}
\end{center}
\end{figure*}

\begin{figure*}
\begin{center}
\vspace{0cm}
\hbox{\epsfxsize=15.0cm \epsfbox{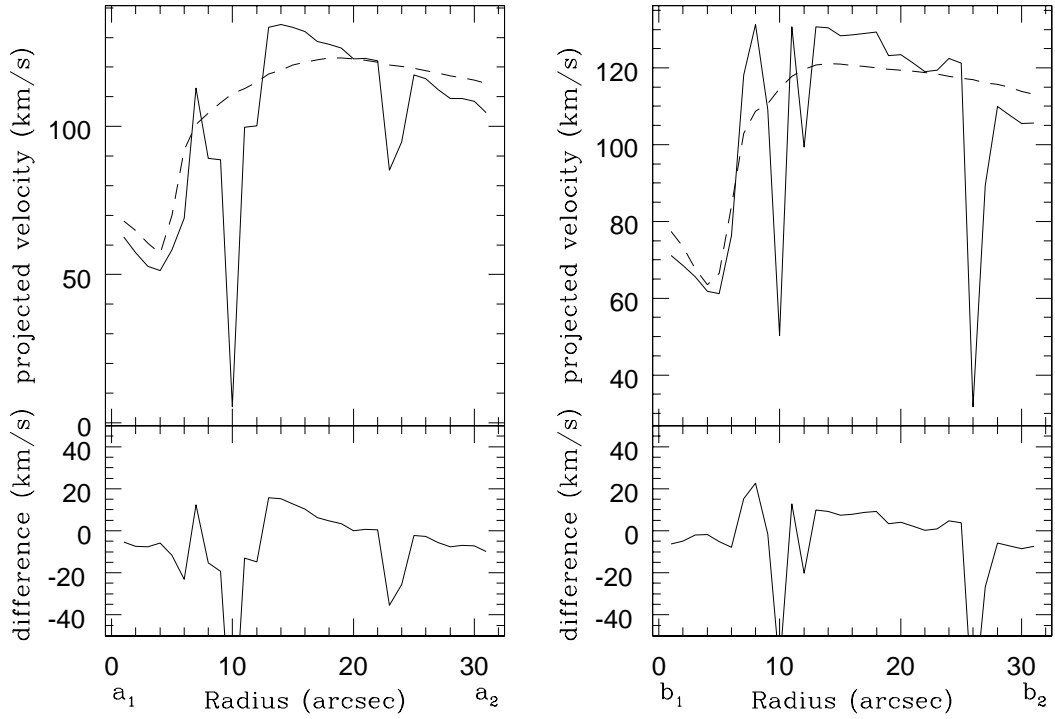}}
\vspace{-3.2cm}
\hspace{-2cm}\epsfxsize=7.0cm \epsfbox{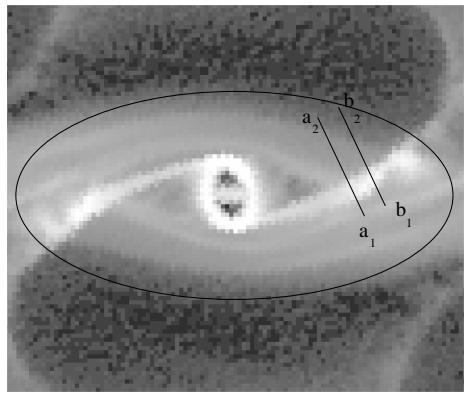}
\caption{This figure shows two cuts across the high velocity gradient region for the deprojected velocity maps of both models. The top panels show the projected velocity. The position corresponds to the velocity going from the leading side towards the trailing side of the shock region in a 30 pixel range ($approx$~23.4~arcsec), with a cut a a distance along the shock of 30 pixels from the centre (left panel) and a distance of 35 pixels from the centre (right panel). The middle panels correspond to the residuals of both line-of-sight velocity distributions. The bottom panels shows the position of the cuts (two parallel black lines) shown in the above panels perpendicular to the shock region on top of the general gas distribution.}
\label{fig:gradiente}
\end{center}
\end{figure*}

After three bar rotations, when the non-axisymmetric component is
fully grown, the gas distribution settles into a steady configuration
until the end of the simulation, six bar rotations. The gas
distribution at the end of the simulation is shown in Fig. 1. The region chosen for a comparison between the two models is the region used by WSW01 in their study  (see
Fig.~\ref{fig:velocity} for the location of the comparison regions). They chose the bar region because is the area where the non-axisymmetric motions are strong and excluded the central region because the velocity gradients are very high and probably unresolved. Their dynamical resolution is worst at the center because there are few cells across the scale of a streamline. Therefore, for a meaningful comparison between the models, we will restrict the comparison to
the regions discussed in WSW01.

Previous work (PA00) in a study of the effect of hydrodynamical and numerical parameters in the simulations of gas flow in barred galaxies, noticed that differences in the sound speed would change the gas distribution morphology and the inflow rate of the gas. The two models under comparison here have similar sound speeds (10\kms for the SPH models, and 8\kms for the grid based models) and therefore, we do not expect differences arising from this effect.

Other parameter that could cause a difference between the two models is the way the non-axisymmetric component is introduced, PA00 noticed that introducing the bar abruptly results in a different morphology when compared to the case where the bar is slowly introduced.  In both models analised in here, the bar is  introduced  slowly, allowing the flow to steadily adjust to the growth of the bar.

The spatial gas resolution in the two codes is very different and difficult to compare. The gas resolution in the SPH is determined by the particle smoothing length, in this case, to ensure good resolution in high density regions, the smoothing length is assigned to  each particle such that their number of neighbouring particles always remains close to a fixed number.  The average gas smoothing length is close to the vertical gravitational resolution, which for NGC~4123 is 20~pc. However, in the central parts, where the particle density is high, the resolution reaches a few parsecs. The resolution of the grid based code is 84.7 pc$^{2}$, using a grid having 256$\times$512 cells. It is interesting to notice that while the gas distribution in the SPH code develops a central ring, associated to the ILR, in the results obtained with the Eulerian code the gas density distribution  does not present  this central feature; although, the underlying potential and the sound speed are similar in both models. Because the mass distribution and the bar pattern speed are similar, the difference in the presence or not of the central ring is possible due to differences between the two codes. PA00 showed the gas distribution for a series of 2-D SPH models with different gas sound speeds and compared them to simulations carried out with a 2-D 
second order flux-splitting code similar to the one used by WSW01. The gas response for a sound speed of 10\kms was similar in both cases, both cases developing a nuclear ring. However, for a sound speed of 15\kms, the gas density response is different  in both codes. The gas density in the SPH code develops  a central x$_{2}$ ring, while the gas distribution in the eulerian code gives no central ring. This should be further investigated using 2-D (or 3-D) consistently. The comparison we carry out  throughout this paper is done outside this central region.   Future high resolution observations of the gas
distribution in the central parts, would show if the velocity fields
and the density distribution generated in our modelling is a good
representation of the central region in NGC~4123 or whether we are
making wrong assumptions on the vertical thickness of the bar or the dynamical approach used here is not appropriate for  the problem of modelling the central regions.  

 
The general characteristics of the velocity fields are the following;
the offset shocks in the bar are roughly east-west and nearly
perpendicular to the bar there is a large velocity gradient. In
Fig.~\ref{fig:velocity}b one can see that, within the comparison region, i.e. the regions associated to the dust lanes the agreement between both models is very good. These regions
are associated to shocks and to regions of high velocity
gradients. 

The velocity field and the corresponding velocity gradients are
similar in both models as shown in Fig.~\ref{fig:velocity}a.  The shape
of the density maxima (accompanied by the high velocity gradients)
puts strong constrains to the potential ~\citep{athanassoula92}. The
angle between the high velocity gradient region and the bar is the same in both
simulations and the shape of the shock regions in both models is
similar, supporting the results from both models. There seems to be a
slight displacement of the peak position of the large gradient region
between both models ($\approx$ 1.5 arcsec), in the SPH the regions of highest density are
broader than in the FS2 code possibly due to the fact that the shock is slightly less well resolved in the SPH model. This difference between SPH and the grid codes has been
already noticed \citet{englmaier}. They obtained broader density peaks
in the shock regions in 2-D SPH modelling compared to 2-D grid
models. 

To compare more quantitatively the two models we have made
cross section profiles of this region of high velocity gradients (see
Fig.~\ref{fig:gradiente}). To facilitate the comparison of the velocity jumps in the shock region, the velocity distribution has been shifted
by two pixels (1.56~ arcsec) because this is the offset found between both models in the peak position of the gradient. The cut positions have been
chosen to be within the comparison region. One has to keep in mind the
statistical nature of SPH when comparing the cuts. This  causes
some particles to have very different values from the general
distribution. The velocity jumps are very similar for both models with a maximum velocity difference between the two
models across this region of 20 \kms. The amplitude of the velocity jump depends on the mass
distribution in the bar region, with thinner and more massive bars
having larger velocity  jumps~\citep{athanassoula92}, and of course, in this case, also on the
position of the cut across the shock. This is what gives rise to the
difference between the right and the left panel in
Fig.~\ref{fig:gradiente}. Overall, the shape and the amplitude of the gas velocity cross section profiles between the two models agree very well.

A change in the parameters such as the pattern speed and the M/L would have caused the velocity field and the shock strength,location and shape to be significantly different  (see ~\cite{athanassoula92,weiner01,perez4}) to the results obtained here.

There is a discrepancy outside the bar region, and the position and
shape of the spiral arms do not agree between the two models. In the
Eulerian code the arms are narrower than in the SPH code; they seem to
circumscribe the arms formed in the SPH code. In any case, the region
of the spiral arms is discarded in the comparison with the observations, both
groups assume a single pattern speed which is probable not the
case for real galaxies.


\section{Conclusions}
\label{conclusions}

The position and strength of the shocks in barred galaxies derived
from dynamical models is robust to the chosen model. Here, we have
compared 2-D to a 3-D code and despite the differences in the two
codes and  the small differences in the mass model, the resulting
velocity and density maps for both models is very similar. However, for similar hydrodynamical parameters, the SPH code develops a central ring, associated to the x$_{2}$ family of orbits while the gas distribution in the FS2 model does not present this central ring. This difference seems to be due to differences between the two codes and not to the difference on the treatment of the vertical forces.  In any case, the position
and the strength of the shocks in both models remains very similar. We can therefore conclude that the methodology used to derive the halo contribution to the potential using dynamical modelling of the gas flow in the bar region of spiral galaxies is robust to the codes used and to wether they are 2 or 3-D codes.

\begin{acknowledgements}
We are very grateful to B. Weiner for providing all the necessary
material that made the comparison possible. We are also grateful to
Reynier Peletier and Almudena Zurita for the careful reading of the manuscript. I. acknowledges financial support from the Netherlands
Organisation for Scientific Research (NWO) Foundation through a VENI grant. This work was also partly funded by the Plan Nacional del Espacio del Ministerio de Educaci\'on y Ciencia espa\~nol.
\end{acknowledgements}

\bibliography{ref}
\bibliographystyle{natbib}

\end{document}